\documentclass[superscriptaddress,twocolumn,showkeys]{revtex4-1}

\usepackage[pdftex]{graphicx}
\usepackage{amssymb, amsfonts, amsmath}
\usepackage{hyperref}

\begin{document}

\title{How molecular knots can pass through each other}
\date{\today}
\author{Benjamin Trefz}
\affiliation{Johannes Gutenberg University Mainz, Department of Physics, Staudingerweg 7, 55128 Mainz, Germany}
\affiliation{Graduate School Material Science in Mainz, Staudinger Weg 9, 55128 Mainz, Germany}
\author{Jonathan Siebert}
\affiliation{Johannes Gutenberg University Mainz, Department of Physics, Staudingerweg 7, 55128 Mainz, Germany}
\author{Peter Virnau}
\affiliation{Johannes Gutenberg University Mainz, Department of Physics, Staudingerweg 7, 55128 Mainz, Germany}

\begin{abstract} 
We propose a mechanism in which two molecular knots pass through each other and 
swap positions along a polymer strand. Associated free energy barriers in our 
simulations only amount to a few $k_{B}T$, which may enable the 
interchange of knots on a single DNA strand.
\end{abstract}


\maketitle 


Ever since Kelvin conjectured atoms to be composed of knots 
in the ether \cite{1}, knots have stimulated the imagination of 
natural scientists and mathematicians alike. In recent years the 
field went through a renaissance and progressed considerably, spurred 
by the realization that topology may not only diversify structure, but 
can also have a profound impact on the function of biological 
macromolecules. Knots in proteins have been reported \cite{2,3,4,5,6,33,7} 
and even created artificially \cite{8}. Topoisomerases can remove \cite{9} 
or create \cite{10} knots in DNA, which may otherwise inhibit 
transcription and replication, and viral DNA is known to be highly 
knotted in the capsid \cite{11,12,13,14,15}. Artificial knots have 
also been tied in single DNA molecules with optical tweezers and 
dynamics have been studied both experimentally and with computer 
simulations \cite{16,17}. Knots are also known to weaken strands, which tend to rupture at the
entrance to the knot \cite{32,34}. Even though most of these examples are not 
knotted in a strict mathematical sense \cite{18}, which only defines 
knots in closed curves, they nevertheless raise fundamental questions 
and challenge our understanding of topics as diverse as DNA ejection \cite{19} 
and protein folding \cite{20}. Knots may also play a role in future 
technological applications, particularly in the advent of DNA nanopore 
sequencing \cite{21}. While the probability of observing a knot in a 
DNA strand of 10 kilo base pairs in good solvent conditions only 
amounts to a few percent \cite{22,23}, knots and even multiple knots 
will become abundant once strand sizes exceed $100000$ base pairs in 
the near future. Part of this problem was recently addressed in a 
simulation study \cite{24}: a single knot will not necessarily jam 
the channel once it arrives at the pore, but may slide along its entrance. 

In the following we would like to elucidate a fascinating and little-known 
property of composite knots: Two knots can diffuse through each other. 
In our simulations we employ a standard bead-spring polymer model \cite{25}, 
which does not allow for bond crossings if local dynamics are applied. 
Furthermore we apply an additional angular potential and tune the stiffness 
of the chain such that in good solvent (high salt) conditions the persistence length of 
DNA is reproduced (for $\kappa = 20 k_{B}T$). Our polymer consists of $250$ monomers 
which corresponds to roughly $1875$ base pairs. Details on our coarse-grained model, 
the mapping onto DNA and determination of knot sizes are given in Materials 
and Methods section. Note that no bias was applied so that our simulations are 
solely driven by thermal fluctuations. As Supporting Information a video of 
one ``tunneling'' event is provided (Video~\ref{SupVid1}).

\begin{figure}[h]
\centering\includegraphics[width=8.7cm, keepaspectratio=true]{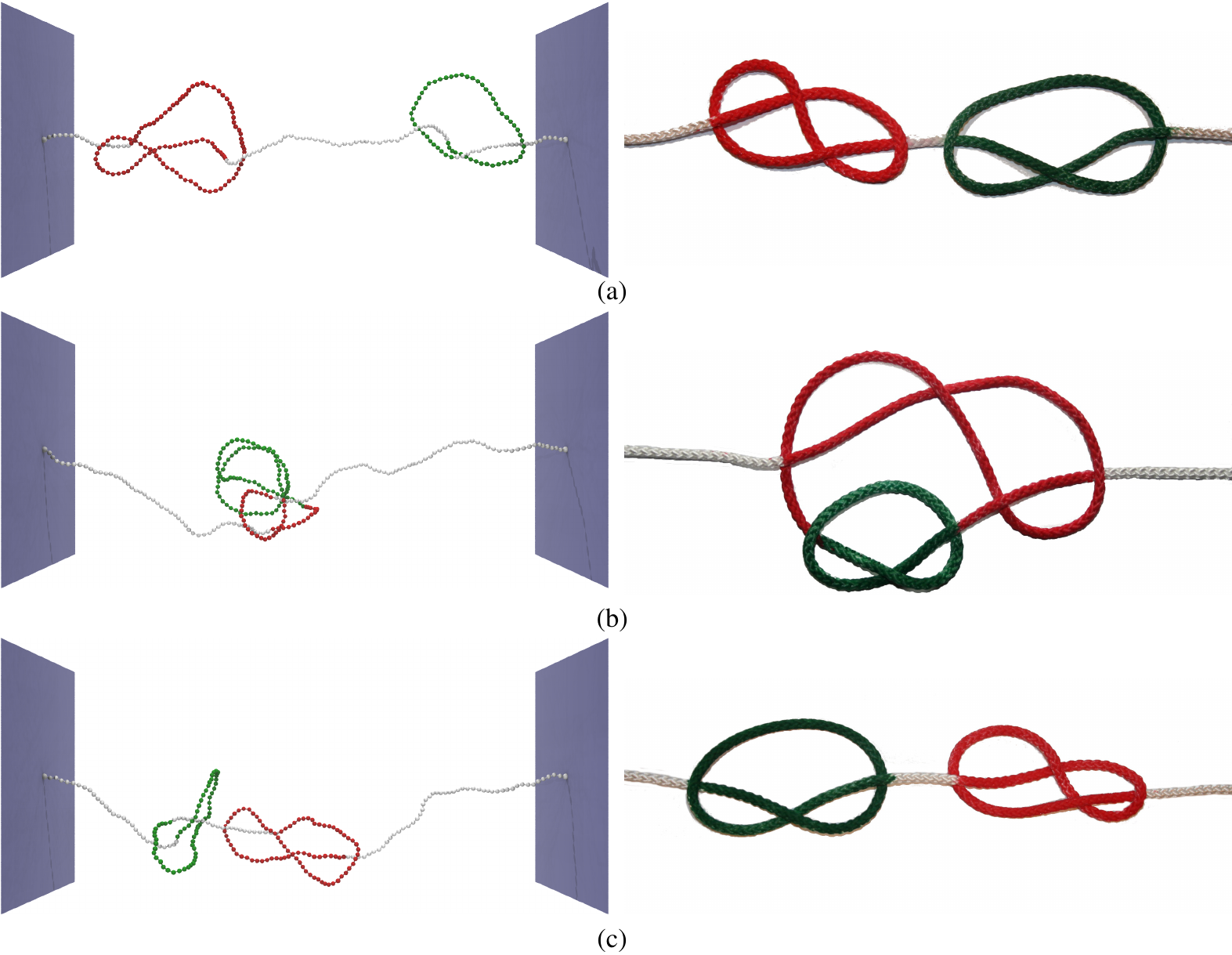}
\caption{%
\label{Fig1}
Left: Snapshot pictures taken before (a), while (b) and after (c) $3_1$ (green) 
and the $4_1$ (red) knots interchange positions along the strand. 
Right: Simplified representation.
}
\end{figure}
In FIG.~\ref{Fig1} we have prepared a starting configuration with a trefoil knot ($3_{1}$) 
on the right hand side (green), which is characterized 
by three non-reducible crossings in a projection onto a plane, and a 
figure-eight knot ($4_{1}$) on the left (red), 
which has four crossings. Both termini are connected to a repulsive wall 
on each side. The distance between walls was chosen to correspond to the 
typical end-to-end distance ($R_{ee}$) of an unentangled polymer strand 
of this size ($d \approx 88 \sigma$). Simulations take place in the NVT 
ensemble (Langevin thermostat) using the CPU version of the HOOMD 
package \cite{26}. Knots are identified using an implementation of the 
Alexander polynomial as described in \cite{27}. The location and the size 
of each knot is determined by successively deleting monomers from both 
ends \cite{28} until we detect first the trefoil or the figure-eight knot 
and finally the unknot or vice-versa. The arithmetic mean of the starting 
and the end monomer of the knot is called its ``center''.

\section{Results}
\begin{figure}[h]
\centering\includegraphics[width=8.7cm, keepaspectratio=true]{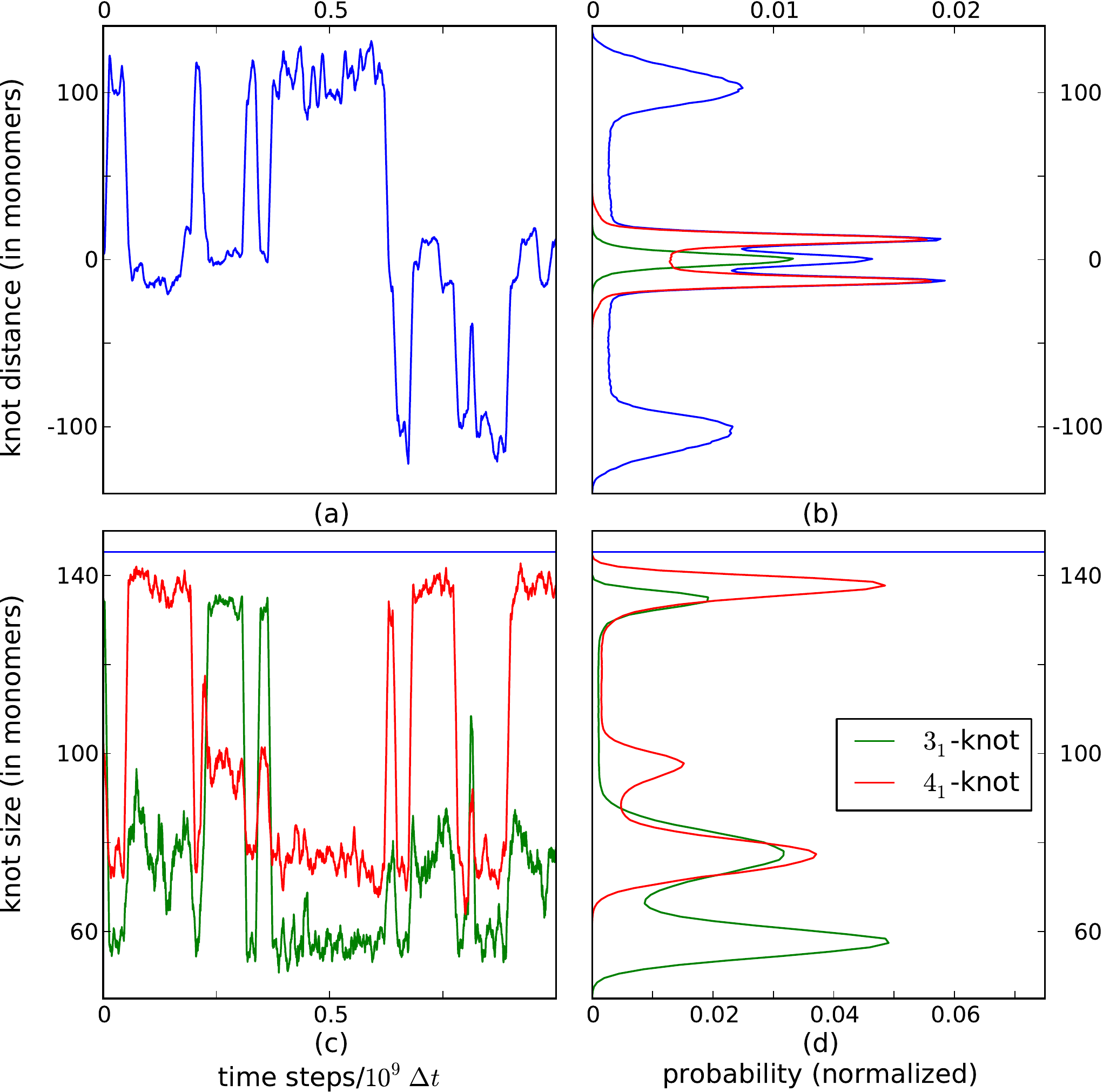}
\caption{%
\label{Fig2}
(a) Distance between the respective ``knot centers'' as a function of simulation time. 
The positions around $+100$ correspond to configurations in which the two knots are separated. 
At $-100$ the knots are also separated, but positions along the strand are 
interchanged. The transition region in which the knots are entangled and pass 
through each other is located around $0$. (b) Corresponding probability profile 
(blue) obtained from (a). Interestingly a triple peak forms in the intertwined state. 
Simulations in which the $4_1$ knot passes through the enlarged $3_1$ knot only 
contribute a single peak (green), while for the opposite situation two peaks 
arise (red). (c) ``Size'' of the trefoil (green) and the figure-eight knot (red) as 
a function of simulation time. The same section was chosen as in (a). 
``Swapping events'' and attempted events are accompanied by a considerable 
enlargement of one of the two knots to around the combined equilibrium size 
of both knots (blue line), while the other knot which diffuses along the big 
knot only grows a bit. (d) Corresponding probability profile obtained from (c). 
The data shown in (a) and (c) is smoothed by applying a running average. For 
details and implications see Materials and Methods. FIG.~\ref{SupFig1} also shows the raw data.
}
\end{figure}
In FIG.~\ref{Fig2}a we follow the location of the knot centers with respect to 
each other, and record their distance (in units of monomers) as a function of 
simulation time. In this framework, the two knots are separated when the two 
centers are around $100$ monomers apart. At $-100$ the knots are also separated, 
but positions along the strand are interchanged. Knots are intertwined when 
centers coincide. As shown in FIG.~\ref{Fig2}a knots may pass through each other 
over and over again via an entangled intermediate state. Can we understand this 
peculiar diffusion mechanism? In FIG.~\ref{Fig2}c (which shows the same section 
as in FIG.~\ref{Fig2}a) we record the size of each knot. When two knots are 
separated the trefoil knot occupies around $60$ monomers, whereas the figure-eight 
knot is slightly larger at around $80$ monomers. In the entangled intermediate 
state one of the knots suddenly expands to a bit less than the combined size of 
the two knots in the separated state, whereas the size of the other knot 
grows only marginally. Intriguingly, it is not always the larger figure-eight 
knot which expands even though its expansion is a bit more likely as can be seen 
in the accumulated histogram in FIG.~\ref{Fig2}d. As the two knot centers more or 
less coincide in the entangled state we conclude that the smaller knot diffuses 
along the strand of the enlarged knot (as depicted in FIG.~\ref{Fig1}b) until the 
two are separated again. They may then either occupy the same positions as before 
or have interchanged positions along the strand. At large stiffness, the probability 
distribution of the intertwined state is split up into a triple peak (FIG.~\ref{Fig2}b) 
which emerges from two separate contributions. If the $4_1$ knot passes through 
the enlarged $3_1$ knot there is only a single peak in the middle (green curve). 
Vice versa, two slightly shifted peaks arise (red curve) due to the symmetry 
of the enlarged $4_1$ knot (compare with FIG.~\ref{Fig1}b). 

\subsection{Topological free energy}
We can also derive an estimate for the ``topological'' free energy barrier which 
needs to be overcome in a ``knot swapping'' event. This barrier essentially
accounts for the obstruction caused by entanglements. In FIG.~\ref{Fig2}b we have 
accumulated data from simulations as shown in FIG.~\ref{Fig2}a to obtain a histogram 
of the time series and a corresponding probability 
distribution. For $\kappa = 20 k_{B}T$ the most likely state is the combined state whereas the separated 
states are metastable. 
\begin{figure}[h]
\centering\includegraphics[width=8.7cm, keepaspectratio=true]{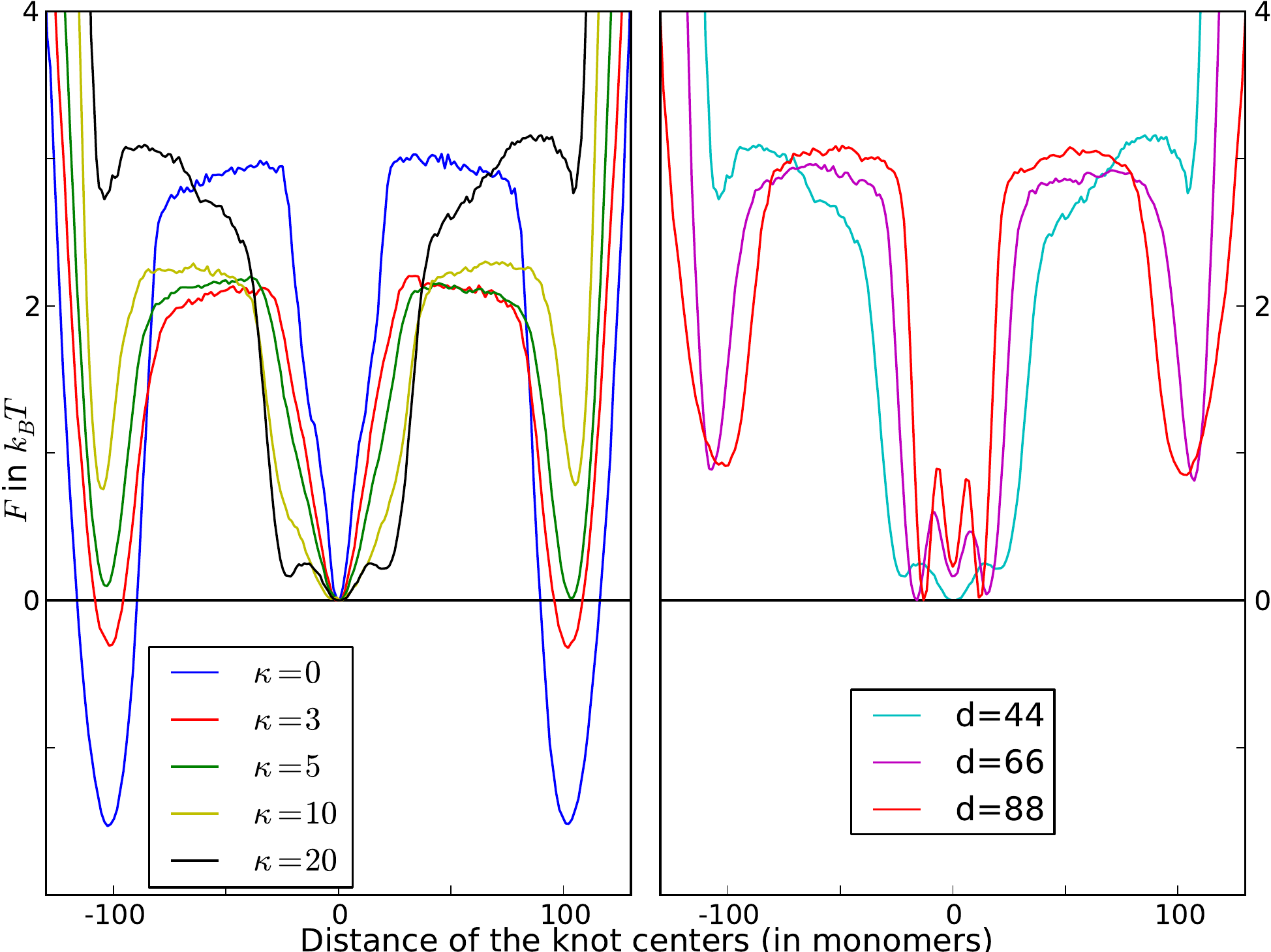}
\caption{%
\label{Fig3}
(a) Free energy profiles derived from probability distributions as shown in 
FIG.~\ref{Fig2}b. The curves correspond to simulations in which the 
stiffness of the chain was modified at a constant wall distance ($d \approx 44 \sigma$) 
to apply more or less tension to the string. While the separated states are 
stable for a flexible chain (blue) the intermediate state becomes stable and 
more likely at higher stiffnesses (green and yellow). The barrier in the 
free energy is also reduced by a higher chain stiffness. Note that smaller values 
for $\kappa$ can also be mapped onto DNA and correspond to larger DNA strands
in physiological and lower salt conditions as detailed in the Materials and 
Methods section. (b) Free energy profiles derived from probability distributions 
as shown in FIG.~\ref{Fig2}b. The three curves correspond to three simulations in 
which the separation of the walls was modified to apply more or less tension to 
the string while the same angular stiffness ($\kappa = 20 k_{B}T$) was used. For all 
wall distances the intertwined state is the most likely state. Note that a 
pronounced triple peak only emerges for large stiffnesses.
}
\end{figure}
From FIG.~\ref{Fig2}b the ``topological'' free energy is derived as 
$F = -k_{B} T \ln(P)$. 
When the separated states are stable (as for flexible chains with $\kappa = 0 k_{B}T$ 
in FIG.~\ref{Fig3}a) the system first needs to overcome a barrier 
${\Delta F_{1} = -k_{B} T} \cdot \allowbreak \ln( P(\text{entrance to intertwined state}) / P(\text{separated state}))$ 
to reach the metastable intertwined state. Then a second barrier 
${\Delta F_{2} = -k_{B} T} \cdot \ln( \allowbreak P(\text{entrance to intertwined state}) \allowbreak/ \allowbreak P(\text{intertwined state}))$
needs to be overcome to finally swap positions or go back to the original state. 
If the intertwined state is stable (as in $\kappa=20 k_{B}T$ in FIG.~\ref{Fig3}b) the 
system needs to overcome $\Delta F_{2}$ to escape into the metastable separated 
state. In all cases the barriers only amount to $2-5 ~ k_{B}T$, which would be 
accessible in experiments. 
Can we alter this barrier? FIG.~\ref{Fig3}a shows free energy profiles from 
simulations with different angular stiffness at the same wall distance. 
While in the case of the lowest stiffness the separated states are more likely, 
the intertwined state is more probable at larger stiffnesses as indicated above. 
FIG.~\ref{Fig3}b 
also shows free energy profiles from simulations in which the walls were placed 
closer together (to $0.5 R_{ee}$ and $0.75 R_{ee}$). While the free energy 
barrier decreases only slightly for $0.75 R_{ee}$, the separated states 
nearly vanish when the two knots are pushed together by the smaller distance 
of the walls (at $0.5 R_{ee}$).

\section{Discussion and Conclusion}
In conclusion, we present a mechanism which allows for two molecular knots 
to diffuse through each other and swap positions along a strand. 
The corresponding free energy barrier in our simulations only amounts to a 
few $k_{B}T$ and should be attainable in experiments similar to \cite{16} 
(with loose composite knots) and potentially in vivo. 
The barrier can be altered by changing the chain stiffness as well as the 
wall distance to make the ``tunneling'' event more or less probable.  
To which extent this peculiar diffusion mechanism might affect DNA behavior 
in nano-manipulation experiments will be investigated in future studies.

\section{Materials and Methods}
\subsection{Model and simulation details}
The model we apply is essentially a discrete variant of the well-known worm-like chain
model (with excluded volume interactions) which has been used extensively to 
characterize mechanical properties of DNA \cite{22, 23, 30, 31}.
We start with a standard bead-spring polymer model from reference \cite{25} which 
does not allow for bond crossings. All beads interact via a cut and shifted 
Lennard-Jones potential (eq.~\ref{eq:lj}). Adjacent monomers interact via the 
finitely extensible nonlinear elastic (FENE) potential (eq.~\ref{eq:fene}). 
Chain stiffness is implemented via a bond angle potential (eq.~\ref{eq:angle}), 
where angle $\theta_{i}$ is measured between the beads $i-1$, $i$ and $i+1$.
For the interaction with the wall we also apply the repulsive part of the 
Lennard-Jones potential (eq.~\ref{eq:wall}), where $d_{i}$ is the orthogonal 
distance from the respective wall to bead $i$. For simplicity we define the 
normal vector of the walls to coincide with the x-axis of our system. 
\begin{align}
    U_{\text{WCA}}(r_{ij}) 	&= \left \{    
				\begin{array}{l l} 
				4 \epsilon \left[ (\sigma / r_{ij})^{12} - (\sigma / r_{ij})^{6} \right] + \epsilon, 	& r_{ij} \le 2^{1/6} \sigma \\
				0, 										& r_{ij}  >  2^{1/6} \sigma 
				\end{array} \right. \label{eq:lj} \\
    U_{\text{fene}}(r_{ij}) 	&= \left \{ 
				\begin{array}{l l} 
				- 0.5  k  R_{0}^{2}  \ln \left[ 1 - (r_{ij} / R_{0})^{2} \right], 		& r_{ij} <   R_{0}	 	    \\
				\infty, 									& r_{ij} \ge R_{0} 
				\end{array} \right. \label{eq:fene} \\
    U_{\text{angle}}(\theta_{i})&= \quad~ \frac{1}{2} ~ \kappa~ ( \theta_{i} - \pi)^2 \label{eq:angle} \\          
    U_{\text{wall}}(d_{i})	&= \left \{ 
				\begin{array}{l l} 
				4 \epsilon \left[ (\sigma / d_{i})^{12} - (\sigma / d_{i})^{6} \right],	& d_{i} \le 2^{1/6} \sigma \\
				0, 									& d_{i}  >  2^{1/6} \sigma 
				\end{array} \right. ,\label{eq:wall}
\end{align}
with $\epsilon = 1 k_{B}T$, $k = 30\ \epsilon / \sigma^{2}$ and $R_{0} =1.5\ \sigma$. 
The two end beads are grafted to the walls and have the same y and z coordinates. 
The simulations are run with the CPU version of HOOMD \cite{26} and use the 
implemented Langevin dynamics thermostat at $T=1$ and $\gamma = 1$. All simulations
take place in a regime where differences of the strain energies along the knots
can be measured but are far away from breaking bonds as seen in \cite{32,34}.
We use a time step of $\Delta t = 0{.}01$ \cite{25} and evaluate the data each 
$10^{5}$ MD-steps.
Each simulation ran for $4\cdot10^9$ MD steps and each parameter set was 
simulated in at least $66$ independent simulations. For $\kappa=10 k_{B}T$, 
$d=44\sigma$, e.g., we have performed 131 independent simulations and observed
973 successful swapping events.

\subsection{Mapping onto DNA}
In the context of knots a similar model was applied in \cite{22} where the  
parameters were obtained from mapping the probability for 
obtaining trefoil knots in the polymer model onto the experimental probability 
observed for a DNA strand of $11.6$ kilo bases as a function of NaCl concentration.
Our model (which is based on this model) has essentially two parameters, which 
can be fitted to mimic real DNA:
The chain stiffness $\kappa$ and the diameter of the chain ($\sigma$ in Lennard-Jones units).
$\sigma$ is taken from \cite{22}. For high salt concentration ($1$ M NaCl), the effective
diameter of the chain is slightly larger then the locus of DNA ($\sigma = 2.5 \text{nm}$). In 
physiological salt conditions ($0.15$ M NaCl) the effective diameter (according to \cite{22})
is somewhat larger ($\sigma = 5 \text{nm}$). For all salt conditions we assume a
persistence length $l_{p}$ of $50 \text{nm}$ or $150$ base pairs.

The relevant energy scale of our model is defined by $\kappa$ in eq. \ref{eq:angle}.
For the discrete worm-like chain model 
\begin{align}
\kappa \approx \frac{l_{p} k_{B} T}{\sigma} \quad. \label{eq:kappa}
\end{align} 
As our model features excluded volume interactions, variable bond lengths 
and angles, we have verified this relation by measuring the persistence length 
(from the decay of the bond angle autocorrelation function) as a function of $\kappa$
in simulations of unbound chains. Hence, for high salt conditions ($\sigma = 2.5 \text{nm}$)
we obtain $\kappa = 20 k_{B}T$. For physiological conditions ($\sigma = 5 \text{nm}$)
$\kappa = 10 k_{B}T$.

From eq.\ref{eq:kappa} we also obtain the persistence length in simulation units. For 
$\kappa = 20 k_{B}T$, $l_{p} \approx 20 \sigma$. Therefore, our chain of $N=250$ 
monomers contains $12.5$ persistence lengths or $12.5 \cdot 150 \text{bp} = 1875$
base pairs. For physiological conditions $l_{p} = 10 \sigma$ and our chain  
corresponds to $25 \cdot 150 \text{bp} = 3750$ base pairs.
(In this calculation we have neglected that the typical distance between adjacent
beads is slightly smaller than $\sigma$.)

To confirm the validity of our model we have undertaken extensive Monte Carlo 
simulations (with fixed bond lengths). We have obtained the probability of observing trefoil knots in a 
$11.6$ kilo base DNA strand in high salt concentration ($1$ M NaCl, $\kappa = 20 k_{B}T$, 
$\sigma = 2.5$ nm, $N=1547$) and physiological salt conditions ($0.15$ M NaCl, 
$\kappa = 10 k_{B}T$, $\sigma = 5.0$ nm, $N=773$). 
In both cases the probability of observing trefoil knots is only slightly smaller 
(just outside the experimental errorbars) than the values for the experimental system \cite{22}.

\subsection{Detection and localization of knots}
To be able to detect knots, the chain has to be closed first. 
This is done by drawing a line outwards and parallel to the walls from the fixed beads.
Then we connect these lines with a large half circle. After the closure we calculate 
the products of  the Alexander polynomials $\Delta_{p}(-1{.}1) = |\Delta(-1.1)\cdot\Delta(-1/1.1)|$ 
as described in reference \cite{27,28}, which yields the composite knot.

\begin{figure}[h]
\centering\includegraphics[width=8.7cm, keepaspectratio=true]{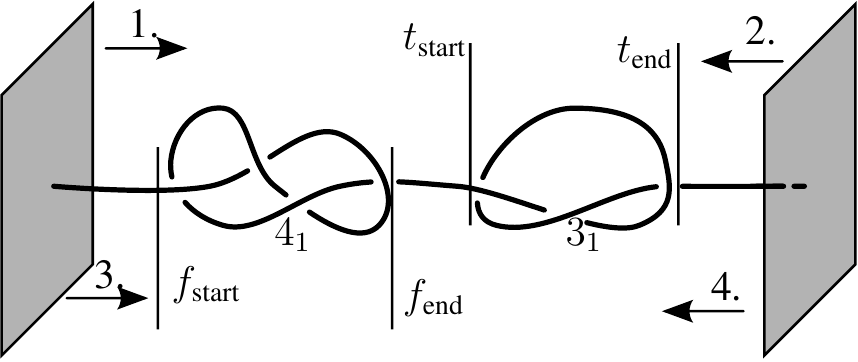}
\caption{%
\label{Fig4}
Schematic drawing of the knot size analysis. 
We start removing beads from the left hand side first ($1.$). 
When the Alexander polynomial yields neither composite nor $3_1$ knot 
the beginning of the $3_1$ knot is reached ($t_{\text{start}}$). 
Restarting this procedure from the right hand side ($2.$) gives us the 
end of the $3_1$ knot ($t_{\text{end}}$). For the $4_1$ knot we start 
from the left hand side again ($3.$) and change the criteria to neither 
composite nor $4_1$ knot which gives us $f_{\text{start}}$. 
Analogously $f_{\text{end}}$ is obtained.}
\end{figure}
For each configuration we confirm that there was no bond-crossing by 
computing $\Delta_{p}(-1.1)$. Now we start to reduce the chain from 
one side by successively removing beads starting with the first bead 
which is not fixed to the wall.
The first remaining bead is connected to the fixed bead on the wall 
and the chain is thereby closed again. 
To determine the starting monomer of one knot, e.g., the $3_1$ knot 
$t_{\text{start}}$, we check after each reduction if the result of 
the Alexander polynomial is still the composite knot or the knot itself.
The end of this knot is determined similarly by starting from the other 
end and applying the same criteria which leads to $t_{\text{end}}$. 
Likewise, we determine the start and the end monomer of the $4_1$ knot $f_{\text{start}}$
and $f_{\text{end}}$.
A scheme of this process is shown in FIG.~\ref{Fig4}. 
With these four values we can calculate both knot centers $m_{1/2}$ on 
the x-axis by using the arithmetic mean. 

\subsection{Data analysis}
The computational determination of knot sizes as described above typically 
results in strongly fluctuating data even if underlying structures are similar. 
This method immanent noise covers up relevant features of the transition such as 
the triple peak in FIG.~\ref{Fig2}b and the slightly increased size of the translocating 
knot in the intertwined state (FIG.~\ref{Fig2}d, compare with FIG.~\ref{SupFig1}). It also 
(artificially) broadens the peaks of the probability distribution at the expense 
of the transition states. For this reason it is not recommended to apply the 
data analysis directly to raw data. Instead, we have chosen to smoothen the data 
by applying a running average over 100 adjacent data points. Note that the 
length of this interval has a minor influence on the barrier height as shown in 
FIG.~\ref{SupFig2}.

\begin{acknowledgments}
P.V. would like to thank M. Kardar for pointing out that two knots on 
a rope may change their position and G. Dietler, C. Micheletti and E. Rawdon 
for helpful discussions.
B.T. and P.V. would like to acknowledge the MAINZ Graduate School of Excellence 
for financial support. 
We would also like to thank F. Rieger for performing the MC simulations in the 
Materials and Methods section and D. Richard for his work on the analysis method 
in the Supporting Information. 
\end{acknowledgments}

\section{Supporting Information}
\renewcommand\thefigure{S\arabic{figure}}
\setcounter{figure}{0}
\begin{figure}[h]
\centering\includegraphics[width=8.7cm, keepaspectratio=true]{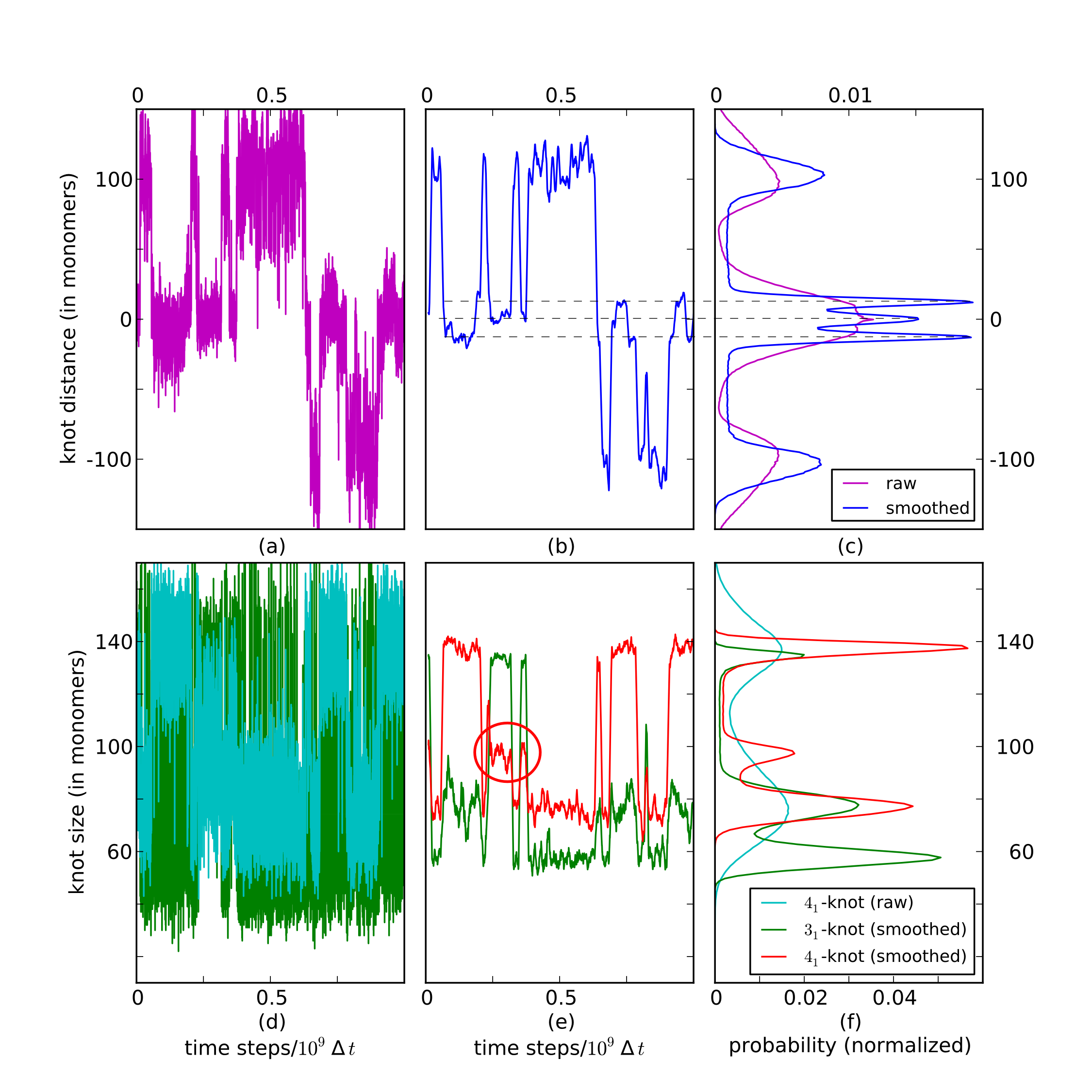}
\caption{%
\label{SupFig1}
Same as FIG.~\ref{Fig2}, but including raw data in panel (a) and (d). 
If our analysis is applied directly to raw data, important 
features such as the occurrence of the triple peak in panel 
(c) and the slightly increased size of the translocating 
(figure-eight) knot in panel (f) are lost in the immanent 
noise of the detection method. The noise also results in an 
artificial broadening of the peaks in panel (c) and (f).
}
\end{figure}

\begin{figure}[h]
\centering\includegraphics[width=8.7cm, keepaspectratio=true]{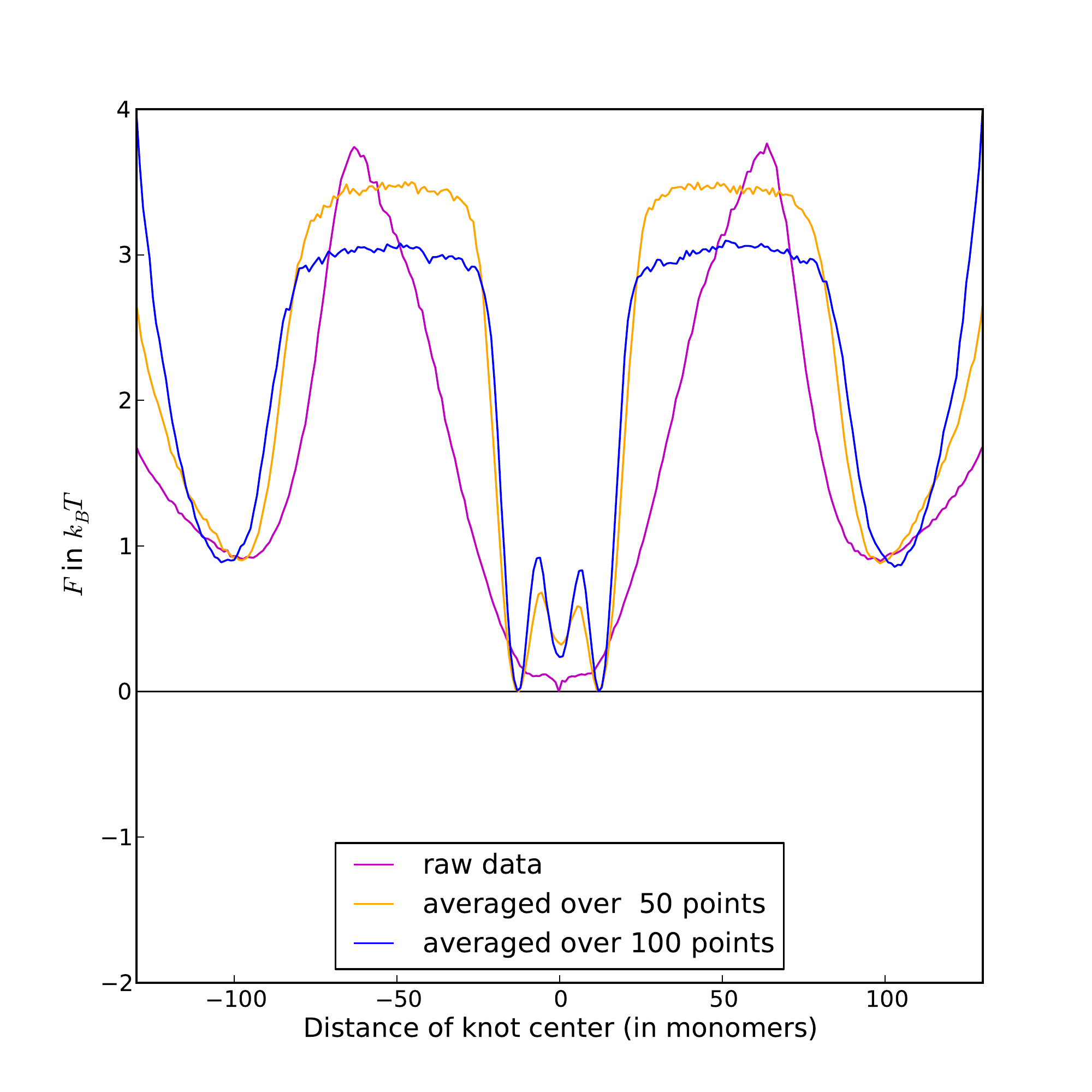}
\caption{%
\label{SupFig2}
Free energy profiles for $\kappa=20k_{B}T$ and $D=88 \sigma$ 
as derived from raw data (magenta) and derived after averaging 
over 50 (orange) and 100 (blue) data points (compare with FIG.~\ref{Fig3}b). 
In the raw data profile the noise artificially broadens the 
minima, which results in a sharper transition state. In addition, 
information on the triple peak is lost. Note that there is a 
small difference in the barrier height ($<0.5 k_{B}T$) if we change 
the interval over which we average from $50$ to $100$ data points.
}
\end{figure}

\begin{video}[h]
\centering\includegraphics[width=8.7cm, keepaspectratio=true]{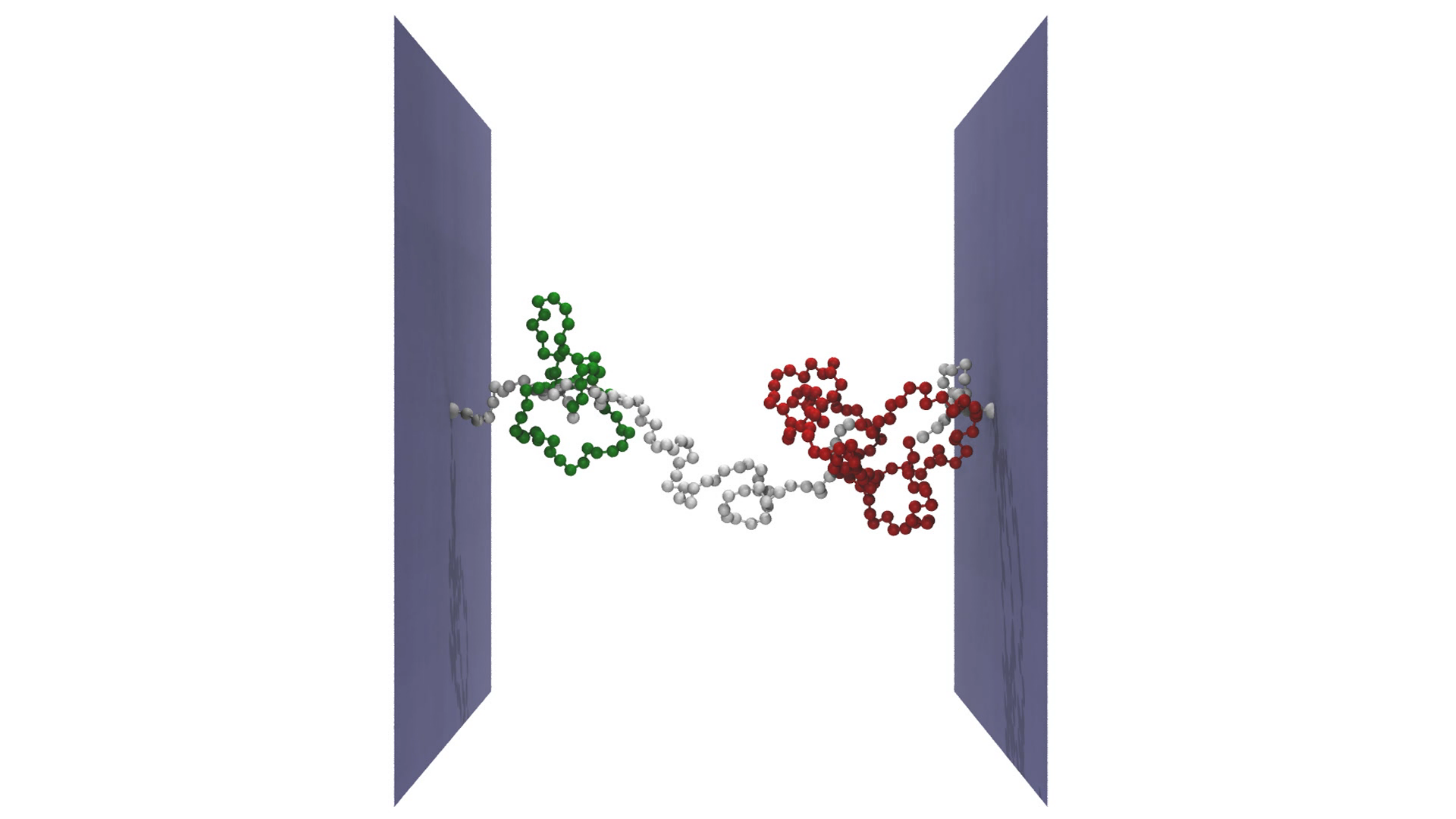}
\setfloatlink{http://www.pnas.org/content/111/22/7948.abstract}
\caption{%
\label{SupVid1}
This video shows a “knot tunneling” event. 
The starting configuration features a trefoil knot 
(colored in green) on the left hand side and a figure-eight 
knot (colored in red) on the right. The trefoil knot 
will diffuse through the enlarged figure-eight knot 
and swap positions with it at the end of the movie.
}
\end{video}

\clearpage{}



\end{document}